\documentclass[12pt]{article}
\usepackage{amsmath,amssymb,epsfig}
\usepackage{cancel}

\makeatletter

\@addtoreset{equation}{section}
\makeatother

\begin{document}

\title{\vbox{
\baselineskip 14pt
\hfill \hbox{\normalsize KUNS-2344}\\
\hfill \hbox{\normalsize RIKEN-MP-23}} \vskip 1.7cm
\bf D term and gaugino masses\\%
in gauge mediation\vskip 0.5cm
}
\author{%
Tatsuo~Azeyanagi$^{1,2}$, \
Tatsuo~Kobayashi$^{1}$,\\ 
Atsushi~Ogasahara$^{1}$\ and 
Koichi~Yoshioka$^{1,3}$
\\*[20pt]
$^1${\it \normalsize
Department of Physics, Kyoto University, Kyoto 606-8502, Japan} \\
$^2${\it \normalsize 
Mathematical Physics Laboratory,
Nishina Center, RIKEN, Saitama 351-0198, Japan} \\
$^3${\it \normalsize 
Department of Physics, Keio University, Kanagawa 223-8522, Japan}
}

\date{}

\maketitle
\thispagestyle{empty}

\begin{abstract}
We systematically study supersymmetry breaking with non-vanishing F
and D terms. We classify the models into two categories and find that
a certain class of models necessarily has runaway behavior of scalar
potential, while the other needs the Fayet-Iliopoulos term to break
supersymmetry. The latter class is useful to have a simple model of
gauge mediation where the vacuum is stable everywhere and the gaugino
mass is generated at the one-loop order.
\end{abstract}

\newpage

\setcounter{footnote}{0}

\section{Introduction}

Supersymmetric extension of the standard model is one of interesting
and promising candidates for physics above the weak scale. Even if
supersymmetry (SUSY) is realized at high-energy regime, it must be
broken below the weak scale, because the superpartners of the standard
model particles have not been observed. It is well known that SUSY
breaking inside our visible sector often leads to phenomenological
problems, for which one usually needs to introduce the hidden sector
for SUSY breaking and its mediation.

Recently, several important properties of SUSY breaking were clarified
in the framework of global SUSY with the canonical K\"ahler potential;
the SUSY-breaking vacuum of F-term scalar potential has pseudo-moduli
or tachyonic direction at tree level. In addition, a superpartner of a
massless fermion remains massless or becomes tachyonic (even) when
SUSY is broken in renormalizable Wess-Zumino models. The latter fact
is called the Komargodski-Shih (KS) lemma in this paper. The general
proof of these properties has recently been given
in~\cite{Ray:2006wk,Sun:2008nh,Komargodski:2009jf}.

The above result has important phenomenological implications,
particularly to the gauge mediation of SUSY breaking. The gauge
mediation is one of interesting mechanisms to mediate SUSY breaking in
the hidden sector to the visible sector at quantum level (for a review,
see~\cite{Giudice:1998bp}). The presence of pseudo-moduli and the
KS lemma imply vanishing gaugino masses at the one-loop order or the
existence of tachyonic direction~\cite{Komargodski:2009jf}. That is
phenomenologically unfavorable since gaugino masses are relatively
light compared with SUSY-breaking scalar masses. Such a problem was
found empirically in a direct gauge mediation
model~\cite{Izawa:1997gs} and a solution has been proposed with
higher-energy local minimum~\cite{Kitano:2006xg}.

It is natural to expect that the result would not hold if one changes
any of the model assumptions, that is, the tree-level F-term scalar
potential of global SUSY Wess-Zumino model with the canonical K\"ahler
form. A possible remedy is to consider the effect of non-canonical
(higher-dimensional) K\"ahler potential~\cite{Nakai:2010th} such as in
supergravity. This change however might spoil the
predictability/calculability of gauge mediation. As for the loop-level
effect, it would be interesting to combine the gauge mediation with
the anomaly mediation~\cite{AM} without losing their merits (see
e.g.~\cite{Pomarol:1999ie}).

Another interesting way out is to include the non-vanishing D-term
scalar potential of additional gauge group. In this paper, we
systematically study the SUSY breaking with F and D-term scalar
potentials of global SUSY models in which the K\"ahler form is
canonical. The models are classified into two categories under the
criterion that the F-flatness conditions have a solution or not. We
then call models are in the first class when there exists no
solution, i.e.\ F-term SUSY breaking. On the other hand, we call
models are in the second class when there exists a solution satisfying
all the F-flatness conditions but it is destabilized by the D-term
contribution. A non-vanishing (charged) F term is generally required
to have a non-vanishing D term, and therefore SUSY-breaking models
with F and D terms belong to each of these two classes.

In the first class, the pseudo-moduli appear in the F-term scalar
potential. As we will show, once the D-term contribution is included,
such a pseudo-moduli direction has a runaway behavior, where the
expectation value runs away to the infinity and the vacuum is
undetermined. In the second class, SUSY breaking can be realized when
and only when a non-vanishing Fayet-Iliopoulos (FI) term is
introduced. We thus find that the second class of models is useful to
avoid the discussion of Komargodski and Shih and to generate one-loop
order gaugino masses in the visible sector through gauge mediation.

This paper is organized as follows. In Section~\ref{sec:Fterm},
we give a brief review on some important properties of SUSY-breaking
models only with the F-term scalar potential. In
Section~\ref{sec:Dterm}, we systematically study generic aspects  
of F and D-term SUSY breaking. The above two classes of models are
investigated in details. In Section~\ref{sec:gaugino_mass}, we
present a simple model of gauge mediation and show that the gaugino
mass is generated on the stable vacuum. Section~\ref{sec:summary} is
devoted to summarizing our results. In Appendix, we summarize the 
stationary conditions for the first class of models discussed in
Section~\ref{sec:Dterm}.

\section{F-term SUSY breaking}
\label{sec:Fterm}

In this section, we briefly review some general features of F-term
SUSY-breaking models, explaining the existence of a pseudo-moduli
direction and the KS lemma as well as their implications to the gauge
mediation.

\subsection{Pseudo-moduli and the KS lemma}

We consider a model with the following general renormalizable
superpotential
\begin{equation}
W=\sum_i f_i\phi_i+\sum_{i,j}\frac{m_{ij}}{2}\phi_i\phi_j 
+\sum_{i,j,k}\frac{\lambda_{ijk}}{6}\phi_i\phi_j\phi_k \ , 
\end{equation}
where $\phi_i$ are chiral superfields and the subscript $i$ denotes
the species of them. The K\"ahler potential has the canonical form;
$K=\sum_i |\phi_i|^2$. Then the scalar potential $V$ receives the
F-term contribution and is written as 
\begin{equation}
V=V_F= \sum_i \overline{W}_{\bar{i}}(\bar\phi)W_i(\phi) \ , 
\end{equation}
where $W_i$ represents the derivative of $W$ with respect to 
$\phi_i$.\footnote{With generalization, we denote the second and third
derivatives of $W$ by
$W_{ij}\equiv \partial_{\phi_i}\partial_{\phi_j}W$ and  
$W_{ijk}\equiv\partial_{\phi_i}\partial_{\phi_j}\partial_{\phi_k}W$,
respectively. A similar notation for the field derivative is used for
the potential $V$ and the D term $D$.} 
Here and hereafter we follow the conventional notation that chiral
superfields and their lowest scalar components are denoted by the same
letters.

Let us suppose that the model has a SUSY-breaking minimum 
at $\phi_i = \phi_i^{(0)}$ where at least one of the F components is
non-vanishing,
\begin{equation}
W_i^{(0)}\equiv W_i(\phi^{(0)})\not=0 \ , 
\end{equation}
for some $i$. Throughout this paper, we use the superscript ${}^{(0)}$
to clarify that it is evaluated at the minimum of the F-term scalar
potential, $\phi_i=\phi_i^{(0)}$. The minimum satisfies the stationary
condition
\begin{equation}
V_j = \sum_i \overline{W}^{(0)}_{\bar{i}}W^{(0)}_{ij}=0 \ .
\label{eq:Wij}
\end{equation}
Since $W^{(0)}_{ij}$ corresponds to the fermion mass matrix, the
condition (\ref{eq:Wij}) implies that there is a massless fermion
along the direction $\overline{W}_{\bar{i}}^{(0)}$ ($\not=0$). This is
nothing but the massless Nambu-Goldstone fermion corresponding to SUSY
breaking.

The boson mass-squared matrix at the minimum is given by
\begin{equation}
M_B^2=\begin{pmatrix}
V_{\bar{i}j} & V_{\bar{i}\bar{j}} \\
V_{ij} & V_{i\bar{j}}
\end{pmatrix}=\begin{pmatrix}
\sum_k \overline{W}_{\bar{i}\bar{k}}^{(0)}W_{kj}^{(0)} &
\sum_k \overline{W}^{(0)}_{\bar{i}\bar{j}\bar{k}}W^{(0)}_k \\[1mm]
\sum_k W^{(0)}_{ijk}\overline{W}^{(0)}_{\bar{k}} & 
\sum_k W_{ik}^{(0)}\overline{W}_{\bar{k}\bar{j}}^{(0)}
\end{pmatrix} ,
\end{equation}
in the basis $(\phi_i,\bar\phi_j)$. Consider a massless fermion
denoted by the eigenvector $v_i$, that is, $W_{ij}^{(0)}v_j =0$. Along
its superpartner direction $(v,\bar{v})$, the boson mass term in the
scalar potential is evaluated as
\begin{equation}
\begin{pmatrix} v\\ \bar{v} \end{pmatrix}^{\!\dagger} \! M_B^2
\begin{pmatrix} v\\ \bar{v} \end{pmatrix}
=\sum_{i,j,k}
v_iW^{(0)}_{ijk}\overline{W}^{(0)}_{\bar{k}}v_j+\text{c.c.} \ ,
\end{equation}
which can be made negative by a phase rotation of the eigenvector
$v$. Hence this must vanish,
\begin{equation}
\sum_{j,k} W^{(0)}_{ijk}\overline{W}^{(0)}_{\bar{k}}v_j =0 \ ,
\label{eq:WWv}
\end{equation}
for a positive semi-definite Hermite $M_B^2$, that is, as long as the
vacuum is stable. In other words, a superpartner of a massless fermion
must be massless or tachyonic (even) when SUSY is 
broken~\cite{Komargodski:2009jf}. In particular, Eq.~(\ref{eq:WWv})
is explicitly written down for the massless Nambu-Goldstone fermion
for SUSY breaking;
\begin{equation}
\sum_{j,k} W^{(0)}_{ijk}
\overline{W}^{(0)}_{\bar{j}}\overline{W}^{(0)}_{\bar{k}}=0 \ . 
\label{eq:Wijk}
\end{equation}
The F-term potential has a tree-level flat direction
(pseudo-moduli) parametrized as
\begin{equation}
\phi_i=\phi^{(0)}_i+z\overline{W}_{\bar{i}}^{(0)} \ , 
\quad z\in \mathbb{C} \ .
\end{equation}
That is easily confirmed by showing that $W_i$ is invariant along this
direction:
\begin{eqnarray}
W_i\big(\phi^{(0)}+z\overline{W}^{(0)}\big) 
= W_i^{(0)} + z\sum_j W_{ij}^{(0)}\overline{W}_{\bar{j}}^{(0)}+
\frac12 z^2 \sum_{j,k}
W_{ijk}^{(0)}\overline{W}_{\bar{j}}^{(0)}\overline{W}_{\bar{k}}^{(0)}
= W_i^{(0)}\ ,
\end{eqnarray}
with the help of Eqs.~(\ref{eq:Wij}) and (\ref{eq:Wijk}).

We have discussed the KS lemma and the presence of pseudo-moduli for
the SUSY-breaking Wess-Zumino models. These two properties hold even
when a model contains higher-order superpotential terms such as in the
generalized O'Raifeartaigh model discussed 
in~\cite{Intriligator:2007py}. The most generic superpotential form of
generalized O'Raifeartaigh models studied in \cite{Komargodski:2009jf} is
\begin{equation}
W = \sum_{i=1}^r X_i f_i(\phi_j) + g(\phi_j) \ .
\label{eq:most-OR}
\end{equation}
The numbers of $X_i$ and $\phi_j$ fields, $r$ and $s$ respectively,
satisfy $r>s$, and $f_i(\phi)$ are generic functions of $\phi$. In
this case, some F components in $X_i$ become nonzero in the
SUSY-breaking vacuum, and the pseudo-moduli direction is therefore a
linear combination of $X_i$. Many concrete examples of SUSY breaking
is classified into this form and share the vacuum property discussed
in this section.

\subsection{Gaugino masses in gauge mediation}

It was shown~\cite{Komargodski:2009jf} that F-term SUSY-breaking
models lead to a phenomenological problem when they are used for the
hidden sector of gauge mediation; vanishing gaugino masses or unstable
hidden sector somewhere in the pseudo-moduli space. To see this, let
us start with the SUSY-breaking Wess-Zumino model in the canonical form
\begin{equation}
W = f X +\sum_{i,j}\frac{1}{2}(\lambda_{ij} X + m_{ij}) 
\phi_i\phi_j +\sum_{i,j,k}\frac{1}{6}\lambda_{ijk}\phi_i\phi_j\phi_k \ . 
\end{equation}
The SUSY-breaking minimum satisfies $W_X \neq 0$ and the scalar
potential has the corresponding pseudo-moduli direction. We assume for 
simplicity that $\det(\lambda X+m)$ is non-vanishing for generic $X$,
that is, massless eigenstates are irrelevant for gauge mediation and
removed from the following discussion. By definition, the determinant
is a polynomial of $X$ and is expanded as
\begin{equation}
 \det(\lambda X + m) = \sum_n c_n(\lambda,m)X^n \,.
\end{equation}
Unless $c_0 \neq 0$ and $c_n = 0$ for $n> 0$, one always finds a
solution for $\det(\lambda X+m)=0$ at some point $X=X_0$ on the
pseudo-moduli direction. This fact means that the mass matrix
$(\lambda X_0 + m)_{ij}$ has a zero eigenvalue and the corresponding
eigenvector $v$ such that $(\lambda X_0+m)v=0$. Then the KS lemma,
Eq.~(\ref{eq:WWv}), is found to imply $\lambda v=mv=0$ if there is no
tachyonic direction anywhere in the pseudo-moduli space. This
contradicts with our assumption that $\det(\lambda X + m)$ is not
identically zero. As a result, $\det(\lambda X+m)$ should not have any
zero, and must be independent of $X$ ($c_n=0$ for $n>0$);
\begin{equation}
\det (\lambda X + m) = c_0 = \text{const.} \ .
\label{eq:const}
\end{equation} 

In gauge mediation, some fields $\phi_M$ play as the messenger of
SUSY breaking which, by definition, have the gauge interaction to the
standard model vector multiplets. The gauge invariance implies the
mass matrix form
\begin{equation}
(\lambda_{ij}X +m_{ij})\phi_i\phi_j =
\begin{pmatrix}
\phi_M \\ \phi'
\end{pmatrix}^T
\begin{pmatrix}
(\lambda X +m)_M & 0 \\
0 & (\lambda X +m)'
\end{pmatrix}
\begin{pmatrix}
\phi_M \\ \phi'
\end{pmatrix} \ .
\end{equation}
The above statement, Eq.~(\ref{eq:const}), holds for each of
sub-matrices, i.e.,
\begin{equation}
\det (\lambda X + m)_M = {\rm const.} \ .
\end{equation}
This fact leads to a problem that the standard-model gaugino masses
are not generated at the leading order of gauge interactions;
\begin{equation}
M_{\tilde g} \propto \frac{\partial}{\partial X}
\log\det(\lambda X + m)_M = 0 \ .
\end{equation}
Thus the vacuum structure of F-term SUSY breaking has the
phenomenological problem (or unstable vacuum) and is not suitable for
gauge mediation. In the next section, we will examine how this
situation is changed when the D-term scalar potential is included.

\section{Including D-term potential}
\label{sec:Dterm}

To circumvent the phenomenologically unfavorable result of F-term SUSY
breaking, we examine in this section how the situation can change with
the D-term potential included. For this purpose, we first present some
useful relations between the F and D components, and then investigate
the vacuum of SUSY-breaking models with F and D terms. Some
discussions of the D-term potential in gauge mediation are given
in~\cite{D-term}.

\subsection{Relations between F and D}

Before proceeding the model classification, we give several useful
relations between F and D components, satisfied identically and only
on the vacuum.

In this paper we discuss the models with superpotential $W$ and U(1)
gauge symmetry. The full scalar potential consists of the F and D terms
\begin{equation}
V = V_F + V_D \ ,
\end{equation}
where the D-term contribution is explicitly given by
\begin{equation}
V_D = \frac{g^2}{2}D^2, \qquad  D = \sum_i q_i|\phi_i|^2 \ .
\end{equation}
Here $g$ denotes the gauge coupling constant and $q_i$ the U(1) charge
of $\phi_i$. The U(1) gauge-invariant Lagrangian generally contains
the linear term of vector superfield (the FI term~\cite{Fayet:1974jb})
with the coefficient $\xi$, and the D component is then modified as
\begin{equation}
D = \xi + \sum_i q_i|\phi_i|^2.
\end{equation}
The formulas given below are valid in the presence of the FI term.

The U(1) transformation acts on $\phi_i$ as 
$\phi_i\to\phi_i+i\varepsilon q_i \phi_i$ where $\varepsilon$ is an
infinitesimal parameter, and the U(1) gauge invariance of the
(super)potential leads to
\begin{equation}
\sum_i W_iq_i\phi_i = 0 \ .
\label{eq:Wqphi} 
\end{equation}
The field derivatives of this identity are also satisfied;
\begin{eqnarray}
\sum_i W_{ij}q_i\phi_i + W_jq_j = 0 \ , \\
\sum_i W_{ijk}q_i\phi_i + W_{jk}(q_j+q_k) = 0 \ ,
\end{eqnarray}
where $j$ and $k$ are not summed over. They are rewritten by using the
derivatives of the D component, 
$D_{\bar{i}}=q_i\phi_i$ and $D_{i\bar{j}}=q_i\delta_{ij}$, as
\begin{eqnarray}
\sum_i W_iD_{\bar{i}} =0 \ ,
\label{eq:WD} \\
\sum_i \big(W_{ij}D_{\bar{i}} +W_iD_{j\bar{i}}\big) = 0 \ , \\
\sum_i \big(W_{ijk}D_{\bar{i}} +W_{ik}D_{j\bar{i}}
+W_{ij}D_{k\bar{i}}\big) = 0 \ .
\end{eqnarray}
Other field identities can be obtained with various contractions, e.g.,
\begin{eqnarray}
\sum_{i,j} W_{ij}D_{\bar{i}}\phi_j = 0 \ , \\
\sum_i q_i|W_i|^2
+\sum_{i,j}W_{ij}\overline{W}_{\bar{i}}D_{\bar{j}} = 0 \ , 
\label{eq:qWW}  \\
\sum_i q_i^2|W_i|^2 -\sum_{i,j,k}D_{\bar{i}}
W_{ij}\overline{W}_{\bar{j}\bar{k}}D_k = 0 \ .
\end{eqnarray}

On the vacuum, the stationary conditions impose additional constraints
between F and D terms:
\begin{equation}
\sum_i W_{ij} \overline W_{\bar i} + g^2 D D_j =0 \ .
\label{eq:Vj}
\end{equation}
By multiplying $\overline{W}_{\bar{j}}$, $\phi_j$, $q_j\phi_j$ and
taking the summation over $j$, we obtain
\begin{eqnarray}
\sum_{i,j}W_{ij}\overline{W}_{\bar{i}}\overline{W}_{\bar{j}} = 0 \ , \\
\sum_{i,j} \phi_iW_{ij}\overline{W}_{\bar{j}} +g^2D(D-\xi) = 0 \ , \\
\sum_i q_i|W_i|^2 - g^2 D \sum_i |D_i|^2 = 0 \ ,
\label{eq:qWWD}
\end{eqnarray}
where we have used the field identities given above. The last relation
means an important fact that the D term gives a non-vanishing
contribution to the scalar potential only if some F terms of
U(1)-charged multiplets are non-vanishing. As seen below, this result
is useful to classify and analyze the models with F and D terms.

\subsection{Classification}

The SUSY vacuum has the vanishing scalar potential, that is, all of F
and D-flatness conditions are simultaneously satisfied.
\begin{alignat}{3}
\text{F flatness :} \quad & W_i = 0 & \quad 
(\text{for}\; {}^{\forall} i) \ , \\
\text{D flatness :} \quad & D \,= \,0 & \ .
\end{alignat}
Thus SUSY-breaking models with F and D terms are classified into two
categories. We call a model is in the first class when there is no
solution satisfying all the F-flatness conditions at the same
time. The second class of models has a solution of the F-flatness
conditions but that is destabilized by the D-term potential, which
causes SUSY breaking. As mentioned above, the D-term effect appears
only if $\sum_iq_i|W_i|^2$ is non-vanishing at the vacuum. Thus the
above two classes are all possible situations leading to SUSY breaking.

For SUSY breaking, only F terms are effective in the first class of
models, and the combination of F and D terms is important in the
second class. We will study the first and second classes of models in
the subsections \ref{subsec:1class} and \ref{subsec:2class},
respectively. The results are summarized in
Table~\ref{table:class}. The third and fourth columns are our main
results in this paper. The scalar potential in the first class of
models turns out to have a runaway direction no matter whether the FI
term is added or not, and thus this class cannot be applied to
phenomenology as it stands. For the second class, SUSY is broken if a
proper FI term is introduced, and it would be useful for model
building such as gauge mediation.

\bigskip

\begin{table}[htbp]
\begin{center}
\begin{tabular}{l|ccc} \hline\hline
      & ~ F-flatness ~ & ~ FI term ~ & ~ FI term ~  \\ 
      &  solution  & $\xi=0$ & $\xi\neq0$ \\ \hline 
First class  & no  & runaway & runaway    \\[1mm]
Second class ~ & yes & SUSY & \cancel{SUSY} \\ \hline
\end{tabular}\medskip
\caption{Classification of SUSY-breaking models with F and D terms.}
\label{table:class}
\end{center}
\end{table}

\subsection{The first class: \ Runaway vacuum}
\label{subsec:1class}

\subsubsection{Non-vanishing D term}

The first class of models belong to the F-term SUSY breaking reviewed
in the previous section. Therefore when only the F-term contribution
is taken into account, the scalar potential has a pseudo-moduli
direction. We first examine whether the situation is changed, 
i.e.\ the pseudo-moduli are uplifted once the D-term contribution is
included. The (field-dependent) bosonic mass-squared matrix is given by
\begin{equation}
M_B^2=\begin{pmatrix}
\sum_k \overline{W}_{\bar{i}\bar{k}} W_{kj}
+g^2(D_{\bar{i}}D_j+DD_{\bar{i}j}) &
\sum_k \overline{W}_{\bar{i}\bar{j}\bar{k}} W_k
+g^2 D_{\bar{i}}D_{\bar{j}}  \\[1mm]
\sum_k W_{ijk} \overline W_{\bar{k}}+ g^2D_iD_j  & 
\!\! \sum_k W_{ik} \overline{W}_{\bar{k}\bar{j}}
+g^2(D_iD_{\bar{j}}+DD_{i\bar{j}})
\end{pmatrix} \ . 
\end{equation}
If the minimum of the full scalar potential corresponds to $D=0$ and 
$W_i\neq0$, the stationary condition (\ref{eq:Vj}) reduces 
to $\sum_i W_{ij} \overline W_{\bar{i}} =0$. Along the massless
direction $\overline{W}_i$, the boson mass term is evaluated as
\begin{equation}
\begin{pmatrix}
\overline{W} \\
W\end{pmatrix}^\dagger\! M_B^2
\begin{pmatrix}
\overline{W} \\
W\end{pmatrix}
=\sum_{i,j,k} W_{ijk}
\overline{W}_{\bar{i}}\overline{W}_{\bar{j}}\overline{W}_{\bar{k}} 
+\text{c.c.} \ , 
\end{equation}
where we have used Eq.~(\ref{eq:WD}). The right-handed side can be
made negative by a phase rotation, which means there is the tachyonic
or pseudo-moduli direction. It is thus found that the situation is not
improved with the D-term potential in this case. Therefore, in the
first class of models, one needs to realize the minimum with $D\neq 0$
to circumvent the phenomenological problem of vanishing gaugino mass.

To realize $D \neq 0$, U(1)-charged scalars must take nonzero
expectation values. Furthermore the previous formula,
Eq.~(\ref{eq:qWWD}), implies that at least one charged F term needs
to be non-vanishing. That is an important condition for viable hidden
sector with this class of SUSY-breaking models. For example, in the
generalized O'Raifeartaigh model (\ref{eq:most-OR}), the fields $X_i$
develop non-vanishing F terms, some of which should have non-vanishing
charges of U(1) symmetry.

The D-term potential $V_D$ is positive semi-definite, and then the
full potential $V$ is generally raised
\begin{equation}
V = V_F + V_D ~\geq~ V_F^{(0)} \ .
\label{eq:Vbound}
\end{equation}
As mentioned before, the superscript ${}^{(0)}$ means that it is
evaluated at the minimum of the F-term potential $V_F$. One naively
expects that the pseudo-moduli direction in $V_F$ is lifted up with a
non-vanishing D term. However, as we will show below, there exists a
runaway direction in the full potential such that $D\rightarrow 0$ and 
$V\rightarrow V_F^{(0)}$.

\subsubsection{Runaway behavior}

\begin{figure}[t]
\begin{center}
\includegraphics[width=7cm]{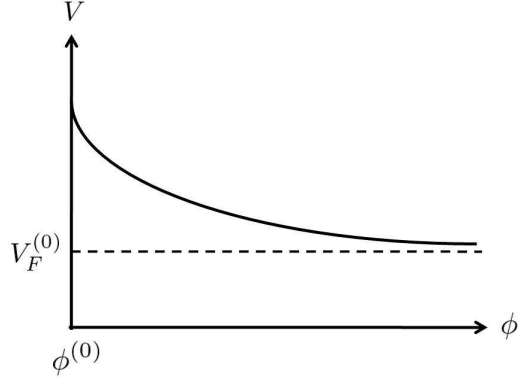}
\caption{The runaway behavior of the full scalar potential along the
direction (\ref{eq:runaway}).}
\label{fig:runaway}
\end{center}
\end{figure}

To show the runaway behavior, let us consider the following direction
\begin{equation}
\phi_i=\phi_i^{(0)}+z_\infty \overline{W}_{\bar{i}}^{(0)}
+\frac{c_i^{(1)}}{\bar{z}_\infty}+\frac{c_i^{(2)}}{\bar{z}_\infty^2} \ ,
\qquad z_\infty \in \mathbb{C}  ,
\label{eq:runaway}
\end{equation}
with a sufficiently large value of $|z_\infty|$. In the limit 
$z_\infty\to\infty$, (\ref{eq:runaway}) is the pseudo-moduli direction
in the potential $V$ without the D term. The F and D terms along this
direction are evaluated as\footnote{We consider the vanishing FI term,
$\xi =0$, for simplicity. The following discussions hold even when a
non-vanishing FI term is included.}
\begin{align}
W_i &= W_i^{(0)}
+\omega^2\sum_{j,k} W_{ijk}^{(0)}\overline{W}_{\bar{j}}^{(0)}c_k^{(1)}
+{\cal O}\Big(\frac{1}{z_\infty}\Big) \ , 
\label{eq:W-runaway}  \\
D &= D^{(0)}
+\sum_i\big(q_iW_i^{(0)}c_i^{(1)}+\text{c.c.}\big)
+{\cal O}\Big(\frac{1}{z_\infty}\Big) \ , 
\label{eq:D-runaway}
\end{align}
where $\omega$ is the phase factor, $\omega=z_\infty/|z_\infty|$. We
have used the field identities (\ref{eq:Wqphi}) and (\ref{eq:qWW}),
and also Eqs.~\eqref{eq:Wij} and (\ref{eq:Wijk}) assuming $V_F^{(0)}$
is the stable minimum of $V_F$. Thus we have $W_i=W_i^{(0)}$ and $D=0$
up to ${\cal O}(1/z_\infty)$, i.e.\ the lower bound (\ref{eq:Vbound})
of the full potential is saturated as $z_\infty\to\infty$ by properly
setting $c_i^{(1)}$ with the fixed values $\phi^{(0)}$'s:
\begin{gather}
\sum_{j,k}W_{ijk}^{(0)}\overline{W}_{\bar{j}}^{(0)}c_k^{(1)} =0 \ , 
\label{eq:WWc} \\
\sum_i q_i \big|\phi_i^{(0)}\big|^2
+\big(q_iW_i^{(0)}c_i^{(1)}+\text{c.c.}\big) = 0 \ .
\label{eq:qWc}
\end{gather}
The illustrative behavior of the potential is given in
Figure~\ref{fig:runaway}.

We give a comment on the roles of the $c_{i}^{(1)}$ and $c_{i}^{(2)}$
terms. It is found from Eqs.~(\ref{eq:W-runaway}) and
(\ref{eq:D-runaway}) that, if $c_{i}^{(1)}=c_{i}^{(2)}=0$, the full
potential has a constant value $V^{(0)}$ along
(\ref{eq:runaway}). However this neither corresponds to the minimum
nor satisfies the stationary condition. The $c^{(1)}$ term drives the
potential to a lower value and the $c^{(2)}$ term stabilizes the
potential. The detail of the potential analysis in the direction
(\ref{eq:runaway}) is given in Appendix.

For the most generic form of the generalized O'Raifeartaigh models
(\ref{eq:most-OR}), the F terms of $X_i$ are non-vanishing, 
$W^{(0)}_{X_i} \neq 0$. On top of that, $X_i$'s appear as the linear
terms in the superpotential and identically satisfy 
$W_{\phi XX}=0$. Therefore the conditions (\ref{eq:WWc}) are
automatically satisfied if the coefficients $c^{(1)}$ are vanishing
for the fields other than $X_i$. The other condition (\ref{eq:qWc})
can be satisfied in the presence of $c^{(1)}_{X_i}\neq0$. Notice that,
as mentioned before, some non-vanishing F components of charged fields
($q_iW_i\neq0$) are necessarily required by a relation between $F$ and
$D$ terms. In the end, the full scalar potential becomes
\begin{equation}
V\to\sum_i\big|W_i^{(0)}\big|^2 \qquad (z_\infty\to\infty) \ ,
\end{equation}
which saturates the lower bound, and the vacuum runs away to infinity.

We conclude that the first class is not relevant for the
phenomenological application to gauge mediation. This class leads to
unstable potential directions in the SUSY-breaking sector or vanishing
gaugino masses at the leading order. The D-term contribution does not
work to cure this problem.

\subsubsection{Example}
\label{subsubsec:runaway_example}

We analyze the runaway behavior in a generalized O'Raifeartaigh model
with the following superpotential
\begin{equation}
W=X_0(f+\lambda \phi_1\phi_2)+m_1X_1\phi_1+m_2X_2\phi_2 \ . 
\end{equation}
The model has a U(1) symmetry, which is gauged, and is relevant to our
discussion of the D-term effect. We take non-vanishing U(1) charges
such that $\phi_1$ and $X_2$ have $+1$, while $\phi_2$ and $X_1$ have
$-1$. The superpotential is the most generic one for this U(1)
symmetry and U(1)$_R$ under which $X_i$ ($i=0,1,2$) have charge 2 and
the others are zero. All the couplings can be taken real positive by
suitable phase rotations up to the U(1) invariance.

Obviously the F-flatness conditions cannot be satisfied
simultaneously, and SUSY is broken. We assume $f\lambda>m_1m_2$ such
that the F terms of charged fields $X_{1,2}$ are non-vanishing. As
seen below, the D term is also non-vanishing in the vacuum when
$m_1\neq m_2$. The minimum of the F-term scalar potential is found
\begin{gather}
\phi_1^{(0)}=-\frac{F}{m_1} \ , \qquad
\phi_2^{(0)}=\frac{F}{m_2} \ , \qquad
X^{(0)}_0=X^{(0)}_1=X^{(0)}_2 =0 \ , 
\end{gather}
where we have defined the combinations of couplings as 
$F\equiv F_0\sqrt{(f\lambda/m_1m_2)-1}$ and 
$F_0\equiv m_1m_2/\lambda$. The expectation values have been made real
by the U(1) rotation. At this minimum, the derivatives of the
superpotential are
\begin{gather}
W^{(0)}_{\phi_1}=W^{(0)}_{\phi_2}=0 \ , \qquad 
W^{(0)}_{X_0}= F_0 \ , \qquad
-W^{(0)}_{X_1}= W^{(0)}_{X_2}= F \ ,
\end{gather}
and the resultant minimum of the F-term potential is given by
$V^{(0)}_F=F_0^2+2F^2$. The F-term potential has the pseudo-moduli
direction along $X_i= z\overline{W}^{(0)}_{\overline{X}_i}$ ($i=0,1,2$).

Let us examine the full scalar potential by adding the D-term
contribution. At the minimum of $V_F$, the D term is non-vanishing,
\begin{equation}
D = \frac{m_2^2-m_1^2}{m_1^2m_2^2}\,F^2 \ ,
\end{equation}
for $m_1 \neq m_2$. This however does not correspond to the full
potential vacuum. Consider a direction parametrized by
\begin{equation}
\phi_i = \phi_i^{(0)} +\frac{c_{\phi_i}^{(2)}}{\bar{z}_\infty^2} \ , 
\qquad X_0=z_\infty W^{(0)}_{X_0} \ , 
\qquad X_i=z_\infty W^{(0)}_{X_i} + 
\frac{c_{X_i}^{(1)}}{\bar{z}_\infty}+
\frac{c_{X_i}^{(2)}}{\bar{z}_\infty^2} \ .
\label{eq:example_runaway}
\end{equation}
It is found that along this direction, the D term approaches to
zero in the limit $z_\infty\to\infty$ keeping the F terms unchanged,
if $c_{X_i}^{(1)}$ satisfy
\begin{equation}
\text{Re}\big(c_{X_1}^{(1)}+c_{X_2}^{(1)}\big) 
+\frac{m_2^2-m_1^2}{2m_1^2m_2^2}\,F =0 \ .
\label{eq:example_condition}
\end{equation}
As a result, the full potential has runaway behavior along the above
direction, i.e.\ $V\to V_F^{(0)}$ as $z_\infty\to\infty$. The
$c^{(2)}$ terms are included to satisfy the stationary conditions as
we discuss in Appendix.

\subsection{The second class: \ Need for the FI term}
\label{subsec:2class}

\subsubsection{Scale transformation}

The definition of the second class is that all F-flatness conditions,
$W_i=0$ for $\!{}^\forall i$, have some solution, but the D-term
contribution destabilizes this minimum of $V_F$, which leads to SUSY
breaking. In this section, we do not consider the case that the F
flatness is satisfied for some infinite values of scalar fields. Such
a possibility would be studied elsewhere~\cite{future-work}.

We have the U(1) symmetry which acts on $\phi_i$ as its phase rotation
with the charge $q_i$. A key ingredient in this section is that the
superpotential is invariant under the following transformation
\begin{equation}
\phi_i \to \tau^{q_i}\phi_i \ ,
\label{eq:cpxU1}
\end{equation}
where $\tau$ is a complex number. That contains a charge-dependent
scale transformation as well as the phase rotation. The F terms
transform as
\begin{equation}
W_i \to \tau^{-q_i}W_i \ .
\label{eq:cpxU1Wi}
\end{equation}
The solution of F-flatness conditions is denoted by $\phi^{(0)}_i$ as
before. The transformation (\ref{eq:cpxU1Wi}) means that, when
$\phi^{(0)}_i$ satisfies the F-flatness conditions,
$\tau^{q_i}\phi^{(0)}_i$ is also a solution. We consider two cases
separately: (i) the F-flatness conditions are satisfied at the origin
of field space, and (ii) the F-flatness requires $\phi^{(0)}_i\neq0$
for some fields. In the second case, the transformation
(\ref{eq:cpxU1}) acts non-trivially on $\phi_i^{(0)}$, which implies
that there is a moduli space spanned by $\tau$.

A simple model for the first case has the superpotential
\begin{equation}
W = m \phi_+ \phi_- \ ,
\label{eq:m+-}
\end{equation}
where $\phi_\pm$ are assumed to have the U(1) charges $\pm 1$. The
solution for $W_{\phi_\pm}=0$ is $\phi_\pm =0$, i.e.\ the origin of
field space. The second case is described, for example, by the
superpotential
\begin{equation}
W = X(f+\lambda \phi_+ \phi_-) \ ,
\label{eq:X+-}
\end{equation}
where $X$ and $\phi_\pm$ have the U(1) charges 0 and $\pm1$,
respectively. The solution for the F-flatness conditions is
\begin{equation}
X=0 \ , \qquad \phi_+ = \tau\phi^{(0)} \ , \qquad 
\phi_- = \tau^{-1}\phi^{(0)} \ ,
\end{equation}
where $\phi^{(0)} =\sqrt{ -{f}/\lambda}$, and $\tau$ is an arbitrary
complex number.

\subsubsection{Including the D term}

Let us consider the effect of D-term potential by gauging the U(1)
symmetry. For the first case, the F-flatness solution
$\phi^{(0)}_i=0$ leads to $D=0$ without the FI term. It is thus found
that a simple SUSY-breaking model for the present purpose is given by
(\ref{eq:m+-}) with a non-vanishing FI term. This model is expected to
satisfy the phenomenological requirement that the gaugino mass is
generated at one-loop order on the vacuum of full scalar potential. We
will discuss the gauge mediation of this type of SUSY breaking in the
next section.

For the second case, the F-flatness conditions have the moduli
direction $\tau^{q_i}\phi^{(0)}_i$ with $\tau$ being a free parameter,
and there the D term is evaluated as
\begin{equation}
D = \xi + \sum_i q_i |\tau|^{2q_i} \big|\phi^{(0)}_i\big|^2 \ ,
\label{eq:D-scale}
\end{equation}
where $\xi$ is the coefficient of the FI term. Suppose that all of
fields with non-vanishing $\phi^{(0)}$ have positive (negative) U(1)
charges. Then, SUSY is found to be broken for a positive (negative)
value of $\xi$. If $\xi=0$, the D-flatness condition requires
$\tau=0$ ($\tau\to\infty$), and the full potential has the
SUSY-preserving vacuum.

For the remaining case that fields with non-vanishing $\phi^{(0)}$
have positive and negative U(1) charges, SUSY is unbroken. This is
because, for a sufficiently small (large) value of $|\tau|$, the D
term (\ref{eq:D-scale}) is dominated by negative (positive) charged
fields and becomes negative (positive), and therefore $D=0$ has a
solution for a finite value of $|\tau|$. For example, the model with
the superpotential (\ref{eq:X+-}) and gauged U(1) symmetry has the D
term in the F-flat direction,
\begin{equation}
D= \xi + (|\tau|^2-|\tau|^{-2})\big|\phi^{(0)}\big|^2 \ .
\end{equation}
The SUSY vacuum is then given by 
$|\phi_\pm|^2=\big(\sqrt{\xi^2+4|\phi^{(0)}|^2}\mp\xi\big)/2$.

To summarize, the second class of SUSY-breaking models should have 
the non-vanishing FI term. We also note that SUSY is unbroken when
both positive and negative-charged scalar fields develop non-vanishing
expectation values.

\section{Gaugino mass generation}
\label{sec:gaugino_mass}

In the previous section, we have classified SUSY-breaking models with
both F and D-term potentials being relevant, and shown that SUSY
breaking is realized in the second class of models with a
non-vanishing FI term. Since the argument by Komargodski and Shih is
not valid in the D-term extension, these models can be applied to the
SUSY-breaking sector of gauge mediation, which generates gaugino
masses at one-loop order.

We consider the model with the following superpotential and the D term 
\begin{equation}
W_{\cancel {\rm SUSY}}=m \phi_+ \phi_- \ , \qquad 
D = \xi + |\phi_+|^2 - |\phi_-|^2 \ .
\end{equation}
The chiral superfields $\phi_\pm$ carry the U(1) charges $\pm 1$. This
model belongs to the second class in our classification. Assuming
$g^2\xi>|m|^2$, we obtain the SUSY-breaking vacuum
\begin{equation}
\langle\phi_+\rangle=0 \ , \qquad 
|\langle\phi_-\rangle|=\sqrt{\xi-\frac{|m|^2}{g^2}} \ ,
\label{eq:model-min}
\end{equation}
and the F and D components are
\begin{equation}
F_{\phi_-}=0 \ , \qquad
F_{\phi_+}=-m^*\langle\bar\phi_-\rangle \ , \qquad 
D=\frac{|m|^2}{g^2} \ .
\label{eq:model-FD}
\end{equation}
This is nothing but the Fayet-Iliopoulos model~\cite{Fayet:1974jb} for
SUSY breaking.

\begin{figure}[htbp]
\begin{center}
\includegraphics[width=6.5cm]{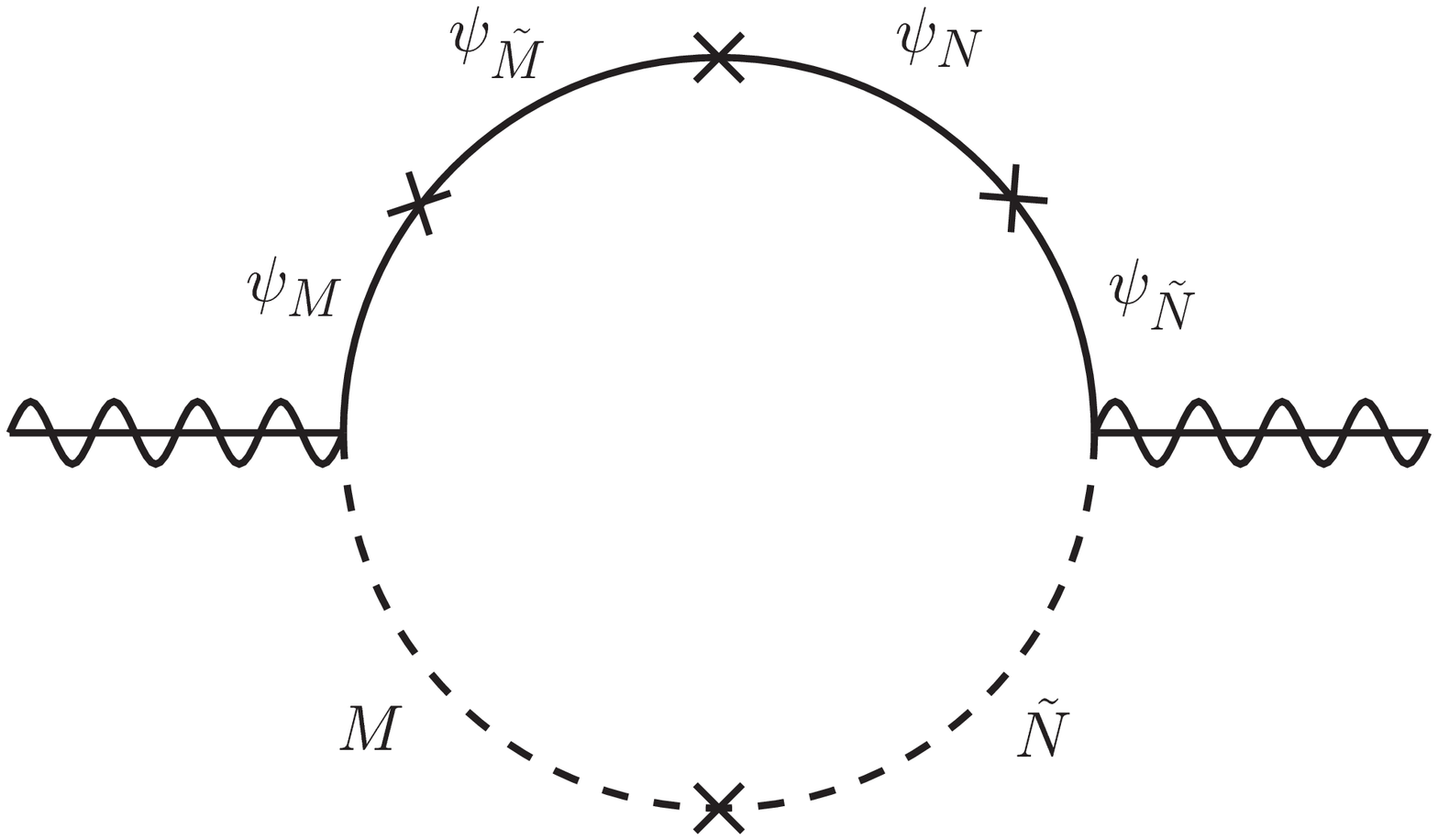}\\[3mm]
\includegraphics[width=6.5cm]{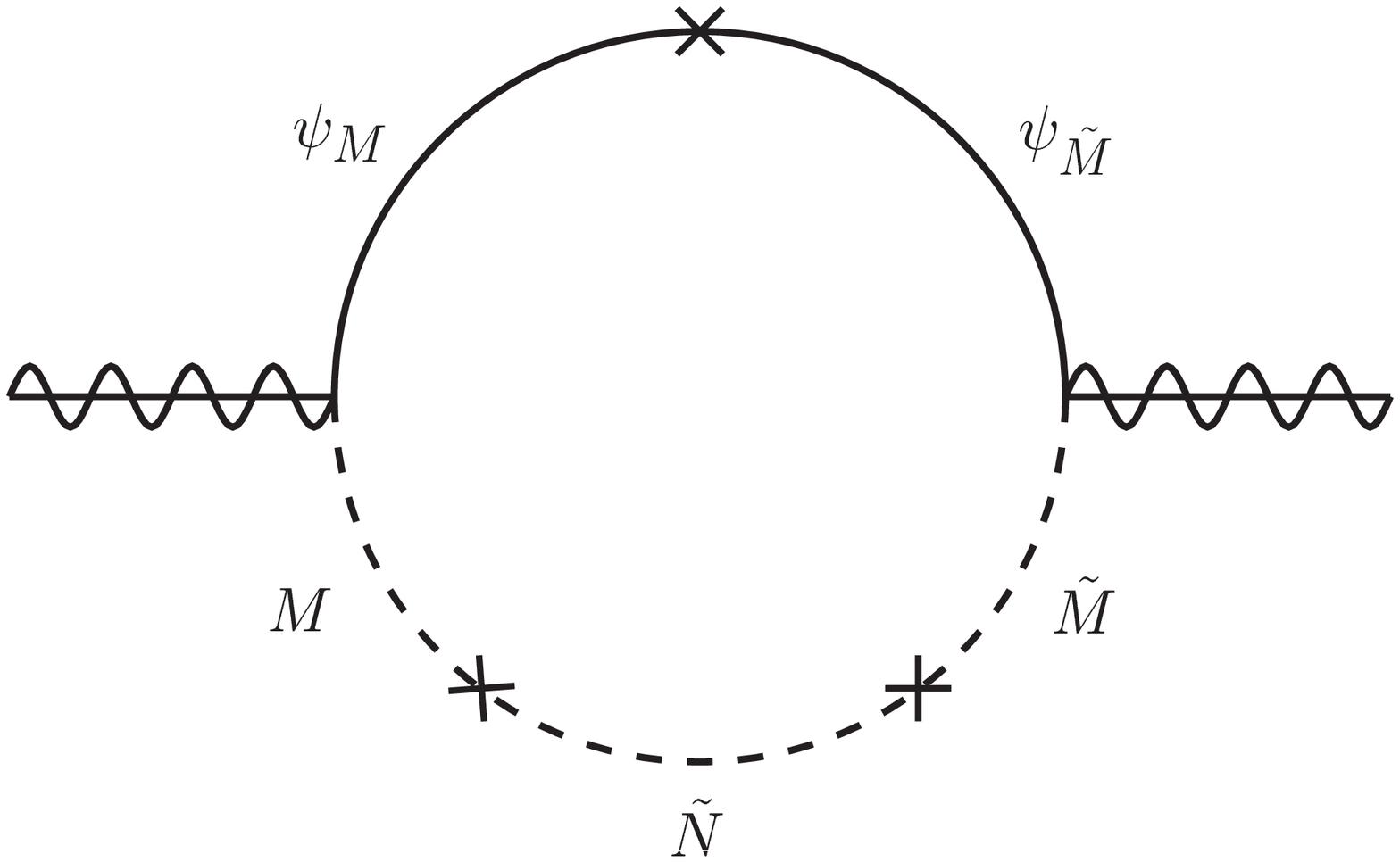}\hspace*{9mm}
\includegraphics[width=6.5cm]{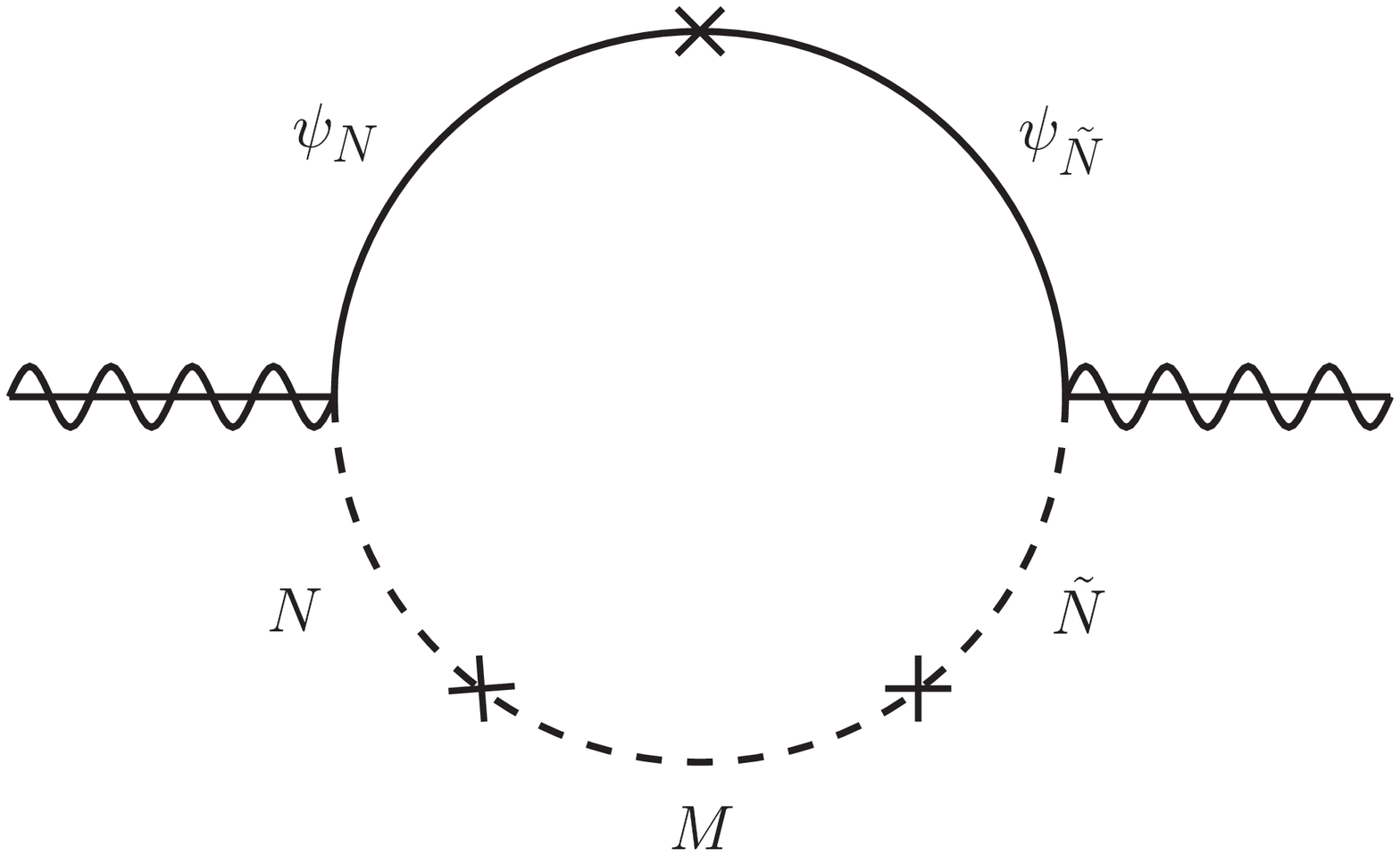}\medskip
\caption{One-loop diagrams for gaugino masses. The symbol $\psi_x$
denotes the corresponding  superpartner of the scalar $x$.}
\label{fig:gaugino}
\end{center}
\end{figure}

We next introduce the messenger sector to mediate the SUSY breaking to
the visible sector via the standard-model gauge interactions. The
messenger chiral multiplets $M$, $\tilde M$, $N$, $\tilde N$ are
charged under the U(1) symmetry and the quantum numbers are shown in
Table~\ref{table:messenger}.
\begin{table}[ht]
\bigskip\begin{center}
\begin{tabular}{c|cccc} \hline\hline
 & ~$M$~ & ~$\tilde{M}$~ & ~$N$~ & ~$\tilde{N}$~ \\ \hline
~U(1)~ & 0 & 0 & 1 & $-1$ \\
$G_\text{SM}$ & $R$ & $R^*$ & $R$ & $R^*$ \\ \hline
\end{tabular}
\caption{The quantum numbers of the messenger multiplets. The 
symbol $R$ means the representation $R$ of the standard model 
group $G_\text{SM}$.}
\label{table:messenger}
\end{center}
\end{table}
As mentioned before, the non-vanishing D term requires charged F terms 
and then implies two pairs of vector-like chiral multiplets as the
minimal set of messengers. We have the superpotential for the
messenger fields,
\begin{equation}
W_\text{mess} = \lambda_+\phi_+M\tilde{N}+\lambda_-\phi_-\tilde{M}N
+m_M^{}M\tilde{M}+m_N^{}N\tilde{N} \ .
\end{equation}
When the messenger mass scales are such that 
$|m_M^{}|,\,|m_N^{}|\gg|m|$, the vacuum (\ref{eq:model-min}) remains
stable and the standard-model gauge group is unbroken. As a result,
the SUSY-breaking F terms are still given by (\ref{eq:model-FD}) and
the bosonic mass-squared matrix for the messengers becomes
\begin{equation}
\begin{pmatrix}
M \\ \bar{\tilde{M}} \\ N \\ \bar{\tilde{N}}
\end{pmatrix}^\dagger\!\!
\begin{pmatrix}
|m_M^{}|^2 & & m_M^* \lambda_- \langle\phi_-\rangle & 
-\lambda_+^* F_{\phi_+}^* \\
& \!|m_M^{}|^2\!+\!|\lambda_-\langle\phi_-\rangle|^2\! & & 
\lambda_- m_N^* \langle\phi_-\rangle  \\
m_M^{}\lambda_-^*\langle\bar{\phi}_-\rangle & & 
\!|m_N^{}|^2\!+\!|m|^2\!+\!|\lambda_-\langle\phi_-\rangle|^2\!\! & \\
-\lambda_+ F_{\phi_+} & \lambda_-^*m_N^{}\langle\bar{\phi}_-\rangle & 
& \!|m_N^{}|^2\!-\!|m|^2
\end{pmatrix}\!
\begin{pmatrix} M \\ \bar{\tilde{M}} \\ N \\ \bar{\tilde{N}}
\end{pmatrix} .
\end{equation}
For a small SUSY breaking $|F_{\phi_+}|\ll|m_{M,N}^{}|^2$, the gaugino
mass $M_{\tilde{g}}$ in the visible sector is generated through the
one-loop diagrams (Figure~\ref{fig:gaugino}) and evaluated as
\begin{equation}
M_{\tilde{g}} = \frac{g_\text{SM}^2}{8\pi^2}
\,T_2(R)\lambda_+\lambda_-
\frac{F_{\phi_+}\langle\phi_-\rangle}{m_M^{}m_N^{}} \ ,
\label{eq:gaugino}
\end{equation}
where $g_\text{SM}^{}$ is the gauge coupling of the standard model
group and $T_2(R)$ denotes the Dynkin index for the messenger
representation $R$. While the diagrams in Figure~2 are expressed in
terms of the mass insertion method, the result of gaugino mass
(\ref{eq:gaugino}) is easily shown to be independent of it and 
exact in the leading 
order.\footnote{The higher-loop effect of U(1) vector multiplet is
negligible with small gauge coupling and/or large breaking scale.}
The result explicitly shows that, in SUSY-breaking models with
both F and D terms, one can indeed circumvent the argument
in~\cite{Komargodski:2009jf} and generate gaugino masses at the
one-loop order of gauge mediation.
\smallskip

\section{Summary}
\label{sec:summary}

In this paper we have systematically studied the SUSY-breaking models
with F and abelian D terms. The analysis has been performed with the
canonical K\"ahler form and the tree-level scalar potential. The
models are classified under the condition whether there exist a
F-flatness solution or not. The first class of models has no solution
for the F-flatness conditions, i.e.\ F-term SUSY breaking, and the
second one has a solution which is destabilized by the D term in the
full potential.

In the first class of models, there exists a pseudo-moduli direction
in the F-term scalar potential as studied
in~\cite{Ray:2006wk,Sun:2008nh,Komargodski:2009jf}. This direction
could be lifted in the full scalar potential by a non-vanishing D term
with finite values of scalar fields. We have however shown that there
is a runaway direction in the full potential and the D term goes to
zero. In the second class of models, SUSY breaking requires a FI term
added with a proper coefficient. Applying a simple model in the latter
class to the gauge mediation, the visible-sector gaugino mass is found
to be generated on the stable vacuum at the one-loop order.

\medskip

\subsection*{Acknowledgment}

The authors thank T.~Okada for useful discussions. T.A.\ is
supported by Japan Society for the Promotion of Science (JSPS) and 
Special Postdoctoral Researchers Program 
at RIKEN\@. T.K.\ and K.Y.\ are supported in part by the
Grant-in-Aid for Scientific Research Nos.~20540266 and 23740187,
and for the Global COE Program "The Next Generation of Physics, Spun
from Universality and Emergence" from the Ministry of Education,
Culture, Sports, Science and Technology of Japan. K.Y.\ is also
supported by Keio Gijuku Academic Development Funds.

\bigskip\bigskip

\appendix

\section{Stationary conditions}

We here examine the stationary conditions of the potential along the
direction (\ref{eq:runaway}). The first derivatives of the F and
D-term potentials with respect to $\bar\phi_i$ are evaluated as
\begin{align}
(V_F)_{\bar{i}} &=\sum_j\Big(\sum_\ell W_{j\ell}^{(0)}c_\ell^{(1)}+
\omega^2\sum_{\ell,m}W_{j\ell m}^{(0)}
\overline{W}_{\bar{\ell}}^{(0)}c_m^{(2)}\Big)
\sum_k \overline{W}_{\bar{i}\bar{j}\bar{k}}^{(0)}W_k^{(0)}
+{\cal O}\Big(\frac{1}{z_\infty}\Big) \ , \\
(V_D)_{\bar{i}} &=g^2\omega q_i\overline{W}_{\bar{i}}^{(0)}
\Big[\sum_j\omega q_j\big(\bar\phi_j^{(0)}c_j^{(1)}
+W_j^{(0)}c_j^{(2)} \big) +\text{c.c.}\Big]
+{\cal O}\Big(\frac{1}{z_\infty}\Big) \ ,
\end{align}
where we have used the field identities (\ref{eq:Wqphi}) and
(\ref{eq:qWW}), and also the minimum conditions of $V_F$, 
i.e.\ Eqs.~(\ref{eq:Wij}) and (\ref{eq:Wijk}). The coefficients
$c^{(1)}_i$ satisfy the equations (\ref{eq:WWc}) and (\ref{eq:qWc})
so that (\ref{eq:runaway}) is a runaway direction. As for the generic
form (\ref{eq:most-OR}), the superpotential is linear in $X_i$ and the
F terms of $\phi_j$ are assumed to be zero, and we find
\begin{align}
(V_F)_{\bar{X}_i} &= 0 \ , \\[2mm]
(V_F)_{\bar\phi_i}\, &= \sum_{\phi_j}
\Big(\sum_{X_\ell} W_{\phi_jX_\ell}^{(0)}c_{X_\ell}^{(1)}+
\omega^2\!\!\sum_{X_\ell,\phi_m} W_{\phi_j\phi_mX_\ell}^{(0)}\!
\overline{W}_{\bar{X}_\ell}^{(0)}c_{\phi_m}^{(2)}\Big)
\sum_{X_k}\overline{W}_{\bar\phi_i\bar\phi_j\bar{X}_k}^{(0)}
W_{X_k}^{(0)} \ , 
\label{eq:VFphij}  \\
(V_D)_{\bar{X}_i} &= g^2\omega q_{X_i}^{}\overline{W}_{\bar{X}_i}^{(0)}
\Big[\sum_{X_j}\omega q_{X_j}^{}\big(\bar{X}_j^{(0)}c_{X_j}^{(1)}
+W_{X_j}^{(0)}c_{X_j}^{(2)} \big) +\text{c.c.}\Big] \ ,
\label{eq:DXj}  \\
(V_D)_{\bar\phi_i}\, &= 0 \ ,
\end{align}
up to ${\cal O}(1/z_\infty)$. Notice that some combination of
$c_{X_i}^{(1)}$'s are determined by the runaway conditions of the
potential and expressed with $\phi^{(0)}$, and generally $c^{(2)}$ are
free parameters. As a result, in the direction (\ref{eq:runaway}), the
stationary conditions for the full potential can be satisfied; 
$V_i=(V_F)_i+(V_D)_i=0+{\cal O}(1/z_\infty)$ with properly
fixed $c^{(2)}$ (and also $c^{(1)}$ consistent with the runaway
behavior). The number of parameters is clearly large enough to have a
solution for the conditions $V_i=0$.

As an example, we discuss the stationary conditions for the model in
Section~\ref{subsubsec:runaway_example}. Along the direction
(\ref{eq:example_runaway}) with a real $z_\infty$, we find the
stationary conditions of the full potential
\begin{align}
V_{\bar{X}_i} &= 2g^2F^2\,
\text{Re}\big(c_{X_1}^{(2)}+c_{X_2}^{(2)}\big) =0 \ ,
\qquad (i=1,2)  \\[1.5mm]
V_{\bar{\phi}_1} &= m_1m_2^2\big(c_{X_2}^{(1)}+
m_1 c_{\phi_1}^{(2)}\big) =0 \ , \\[1mm]
V_{\bar{\phi}_2} &= m_1^2m_2\big(c_{X_1}^{(1)}+
m_2 c_{\phi_2}^{(2)}\big) =0 \ ,
\end{align}
by explicitly writing down Eqs.~(\ref{eq:VFphij}) and (\ref{eq:DXj}). 
The condition for the charge-neutral field $X_0$ is trivial. Together
with the runaway condition (\ref{eq:example_condition}), they give 6
conditions among 12 real parameters, $c^{(1)}_{X_i}$, $c^{(2)}_{X_i}$,
and $c^{(2)}_{\phi_i}$. We thus find runaway solutions, an example of
which is described by
\begin{align}
\phi_1 &= -\frac{F}{m_1} \ , \\
\phi_2 &= +\frac{F}{m_2}
+\frac{1}{z_\infty^2}\,\frac{m_2^2-m_1^2}{2m_1^2m_2^3}\,F \ , \\[1mm]
X_0 &= z_\infty F_0 \ , \\[.5mm]
X_1 &= -z_\infty F
-\frac{1}{z_\infty}\,\frac{m_2^2-m_1^2}{2m_1^2m_2^2}\,F \ , \\[.5mm]
X_2 &= +z_\infty F \ .
\end{align}

\end{document}